\documentclass[11pt,oneside]{article}
\usepackage{amssymb,bm}
\usepackage{amsmath}
\usepackage{amsfonts}
\usepackage{verbatim}
\usepackage{psfrag,graphicx}
\usepackage{amsmath,natbib}
\usepackage{setspace}
\usepackage{subfigure}
\usepackage[left = 1.2in, top = 4in]{geometry}
\usepackage[page]{appendix}

\oddsidemargin %-0.2in
\evensidemargin %-0.2in
\bibpunct{(}{)}{;}{a}{}{,}

 \newtheorem{lemma}{Lema}[section]

\newcommand{\MCMC}{Markov chain Monte Carlo}
\newcommand{\MH}{Metropolis-Hastings}

\newcommand{\jmo}{j-1}
\newcommand{\dof}{degrees of freedom}
\newcommand{\cdf}{cumulative distribution function}
\newcommand{\cdfs}{cumulative distribution functions}
\newcommand{\Ehat}{\widehat E}
\newcommand{\varhat}{\widehat {\text{var}}}
\newcommand{\phat}{\widehat p}
\newcommand{\ect}{equivalent computing time}
\newcommand{\bfy}{\mathbf{y}}
\newcommand{\bfx}{\mathbf{x}}
\newcommand{\bfz}{\mathbf{z}}

\newcommand{\bfyi}{\mathbf{y}_i}
\newcommand{\bfxi}{\mathbf{x}_i}
\newcommand{\bfbeta}{\bm{\beta}}
\newcommand{\bfmu}{\bm {\mu } }
\newcommand{\bftheta}{\bm{ \theta }}
\newcommand{\bfSigma}{\bm{\Sigma}}

\newcommand{\MATLAB}{MATLAB}
\newcommand{\IG}{\text{IG}}
\title{\textbf{
 \\A copula based approach to adaptive sampling}\vspace{0.5cm}}
 \author{Ralph S. Silva\\{\small School of Economics}\\
 {\small University of New South Wales}\\{\small r.silva@unsw.edu.au}
 \and Robert Kohn
 \\{\small School of Economics}\\{\small University of New South Wales}\\
 {\small r.kohn@unsw.edu.au}\\\vspace{0.75cm}
 \and Paolo Giordani\\{\small Research Department}\\{\small Sveriges Riksbank}\\
 {\small paolo.giordani@riksbank.se}
 \and Xiuyan Mun\\{\small School of Banking and Finance}\\
 {\small University of New South Wales}\\{\small ziki.mun@gmail.com}}

 \date{\vspace{0.5cm}{December 31 2008}}
 
\begin{document}
 \maketitle
 \thispagestyle{empty}
 \begin{abstract}
Our article is concerned with adaptive sampling schemes for Bayesian
inference that update the proposal densities using previous
iterates. We introduce a copula based proposal density which is made more efficient
by combining it with antithetic variable sampling. We compare the copula based proposal
to an adaptive proposal density based on a multivariate mixture
of normals and an adaptive random walk Metropolis proposal. We also
introduce a refinement of the random walk proposal which performs
better for multimodal target distributions.
We compare the sampling schemes using
challenging but realistic models and priors applied to real data
examples. The results show that for the examples studied, the
adaptive independent \MH{} proposals are much more efficient than
the adaptive random walk proposals and that in general the
copula based proposal has the best acceptance rates and lowest
inefficiencies.

 \end{abstract}

% \linenumbers
 \newpage
 \setstretch{2.0}
%----------------------------------------------------------------------------------
% \section{Introduction}\label{section:introd}
% \section{Adaptive sampling algorithms}\label{section:adap:samp}
% \subsection{Adaptive random walk Metropolis}\label{SS: arwm}
% \subsection{Adaptive independent Metropolis-Hastings}\label{SS: aimh}
%\subsection{A mixture of a $t$ copula and a multivariate $t$ proposal %distribution} \label{SS: t copula and multi t}
%\subsection{A $t$ copula based proposal density} \label{SS: copula based proposal}
%\section{Applications}\label{section:applic}
\section{Introduction}\label{section:introd}
%Adaptive sampling algorithms are based on Markov chain Monte Carlo
% \citep{metr:rose:rose:tell:tell:1953,hast:1970} methods.
Bayesian inference using \MCMC{} simulation methods is used extensively in
statistical applications. In this approach, the parameters are generated from
a proposal distribution, or several such proposal distributions, with the
generated proposals accepted or rejected using the \MH{} method; see for example
\cite{tierney1994}.

In adaptive sampling the  parameters of the proposal distribution
are tuned by using previous draws. Our article deals with
diminishing adaptation schemes, which means that the difference
between successive proposals converges to zero. In practice, this
usually means that the proposals themselves eventually do not
change. Important theoretical and practical contributions to
diminishing adaptation sampling were made by
\cite{holden98}, %Haario et al.~(2001)
\cite{haario01}, %Andrieu and Robert~(2001)
\cite{gaas:2003},
\cite{andrieu01}, %Andrieu et al.~(2005)
\cite{andrieu05}, %Andrieu and Moulines (2006)~
\cite{andrieu06},
%Roberts and Rosenthal~(2007)
\cite{roberts07} and
%Roberts and Rosenthal~(2006)
\cite{roberts06_applied}.
The adaptive random walk Metropolis method was proposed by %Haario et al.~(2001)
\cite{haario01} with
further contributions by %Atchad\'{e} and Rosenthal~(2005)
\cite{atchade05}, %Andrieu et al.~(2005)
\cite{andrieu05} and
%Roberts and Rosenthal~(2006)
\cite{roberts06_applied}. %Giordani and Kohn~(2008)
\cite{gior:kohn:2008} propose an adaptive independent
\MH{} method with a mixture of normals proposal which is estimated using a
clustering algorithm.

Although there is now a body of theory justifying the use of
adaptive sampling, the construction of interesting adaptive samplers
and their empirical performance on real examples has received less
attention. Our article aims to fill this gap by introducing a
%adaptive sampling scheme using a
$t$-copula based proposal density. An antithetic version of this proposal is also studied
and is shown to increase efficiency when the acceptance rate is above 70\%.
We also refine the adaptive \MH{}
proposal in \cite{roberts06_applied} by adding a heavy tailed
component to allow the sampling scheme to traverse multiple modes
more easily. As well as being of interest in its own right, in some of the examples
we have also used this refined sampler to initialize the adaptive
independent \MH{} schemes. We study the performance of the above
adaptive proposals, as well as the adaptive mixture of normals
proposal of \cite{gior:kohn:2008}, for a number of models and priors
using real data. The models and priors produce challenging but
realistic posterior target distributions.

\cite{silva:kohn:giordani:mun2008} is a longer version of our
article that considers some alternative versions of our algorithms and includes more details and examples.

\section{Adaptive sampling algorithms}\label{section:adap:samp}
Suppose that $\pi(x)$ is the target density from which we wish to
generate a sample of observations, but that it is computationally
difficult to do so directly. One way of generating the sample is to
use the \MH{} method, which is now described. Suppose that given some
initial $x_0$ we have generated the $j-1$ iterates $x_1, \dots,
x_{\jmo}$. We generate $x_{j}$ from the proposal density $q_{j}(x;z)
$ which may also depend on some other value of $x$ which we call
$z$. Let $x_{j}^p$ be the proposed value of $x_{j}$ generated from
$q_{j}(x_{\jmo};x) $. Then we take $x_{j} = x_{j}^p$ with
probability
\begin{align} \label{e:adaptive accep prob}
\alpha(x_{\jmo};x_{j}^p) & =
\min \biggl \{1,
\frac{\pi(x_{j}^p)}{\pi(x_{\jmo})}
\frac{q_{j}(x_{\jmo};x_j^p)}{q_{j}(x_j^p;x_{\jmo})}
\biggr \} \ ,
\end{align}
and take $x_j = x_{\jmo}$ otherwise. If $q_j(z;x)$ does not depend on $j$, then
under appropriate regularity conditions we can show that the sequence of iterates
$x_j$ converges to draws from the target density $\pi(x)$. See
\cite{tierney1994} for details.

In adaptive sampling the parameters of $q_j(z;x)$ are estimated from the iterates
$x_1, \dots, x_{j-2}$. Under appropriate
regularity conditions  the sequence of iterates $x_j, j \geq 1$,
converges to draws from the target distribution $\pi(x)$. See
\cite{roberts07}, \cite{roberts06_applied} and \cite{gior:kohn:2008}.

We now describe the adaptive sampling schemes studied in the paper.
% Adaptive Random Walk Metropolis
%--------------------------------------------------------------------------------------------------
 \subsection{Adaptive random walk Metropolis}\label{SS: arwm}
The adaptive random walk Metropolis proposal of %Roberts and Rosenthal~(2006)
\cite{roberts06_applied} is
\begin{align} \label{e:arwm proposal}
q_n(z;x)& =\omega_{1n}\phi_d (x; z, \kappa_1\Sigma_1) + \omega_{2n}\phi_d (x; z, \kappa_2\Sigma_{2n})
\end{align}
where $d$ is the dimension of $x$ and $\phi_d(x; z, \Sigma)$ is a multivariate $d$
dimensional normal density in $x$ with mean $z$ and covariance matrix $\Sigma$. In
\eqref{e:arwm proposal}, $\omega_{1n} = 1$ for $n \leq n_0$, with $n_0$
representing the
initial iterations, $\omega_{1n} = 0.05$ for $n > n_0$ with
$\omega_{2n} = 1 -  \omega_{1n}$; $\kappa_1 = 0.1^2/d, \kappa_2 = 2.38^2/d,
\Sigma_1 $ is a constant covariance matrix,
which is taken as the identity matrix by %Roberts and Rosenthal~(2006)
\cite{roberts06_applied}
 but can be based on the Laplace approximation or some other estimate. The matrix
 $\Sigma_{2n}$ is the sample covariance matrix of the first $n-1$ iterates.
The scalar $\kappa_1$ is meant to achieve a high acceptance rate by moving the
sampler locally, while the scalar $\kappa_2$ is considered to be optimal %(Roberts et al., 1997)
\citep{robe:gelm:gilk:1997} for a random walk proposal when the target is
multivariate normal.
We note that the acceptance probability \eqref{e:adaptive accep prob} for the
adaptive random walk Metropolis simplifies to
\begin{align} \label{e:arwm accep prob}
\alpha(x,z) = \min\left(1,\frac{\pi(z)}{\pi(x)}\right).
\end{align}

We refine the two component random walk Metropolis proposal in
\eqref{e:arwm proposal} by adding a third component with
$\Sigma_{3n}=\Sigma_{2n}$ and with $\kappa_3 = 25 \gg \kappa_1,
\kappa_2$. We take $\omega_{3n} = 0$ if $n \leq n_0 $, $\omega_{3n}
= 0.05 $ for $n > n_0$ and $\omega_{2n} = 1- \omega_{1n} -
\omega_{3n}$. Alternatively,  the  third component can be a
multivariate $t$ distribution with small degrees of freedom. We
refer to this proposal as the three component adaptive random walk.
The purpose of the heavier tailed third component is to allow the
sampler to explore the state space more effectively by making it
easier to leave local modes.

To illustrate this issue we consider the performance of the two and
three component adaptive random walk samplers when the target
distribution is a two component and five dimensional multivariate
mixture of normals. Each component in the  target has equal
probability, the first component has mean vector
$\mu_1=(-3,\ldots,-3)^T$ and the second component has mean vector
$\mu_2=(3,\ldots,3)^T$. Both components have identity covariance
matrices. For the three component adaptive random walk we choose
$\kappa_3=4^2$. The starting value is $x_0=\mu_1$ for both adaptive
random walk samplers.

\begin{figure}[!t]
\centering
\subfigure[RWM:wrong marginal]
{\includegraphics[scale=0.25,angle=-90]{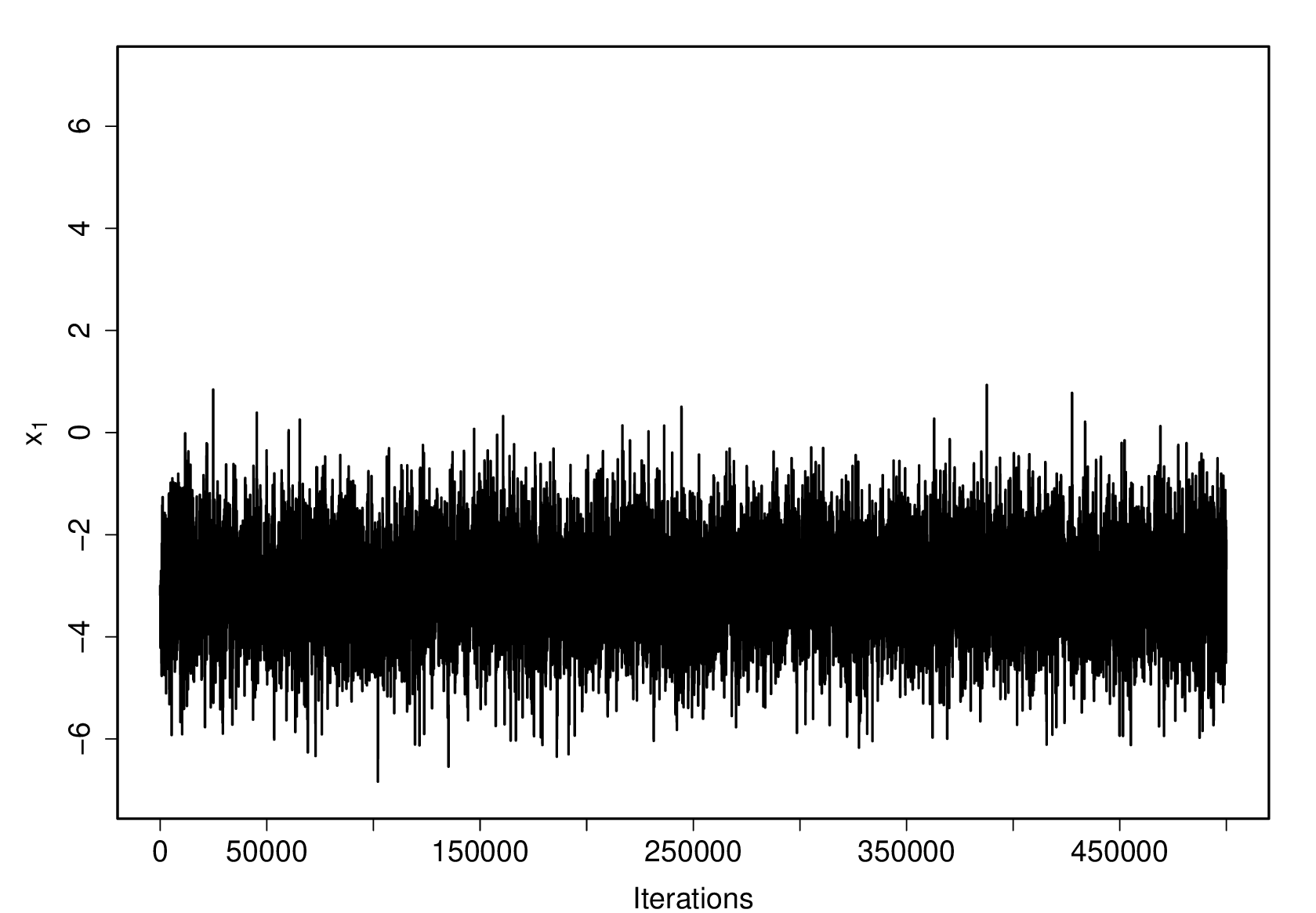}}
\subfigure[RWM3C:correct marginal]{\includegraphics[scale=0.25,angle=-90]{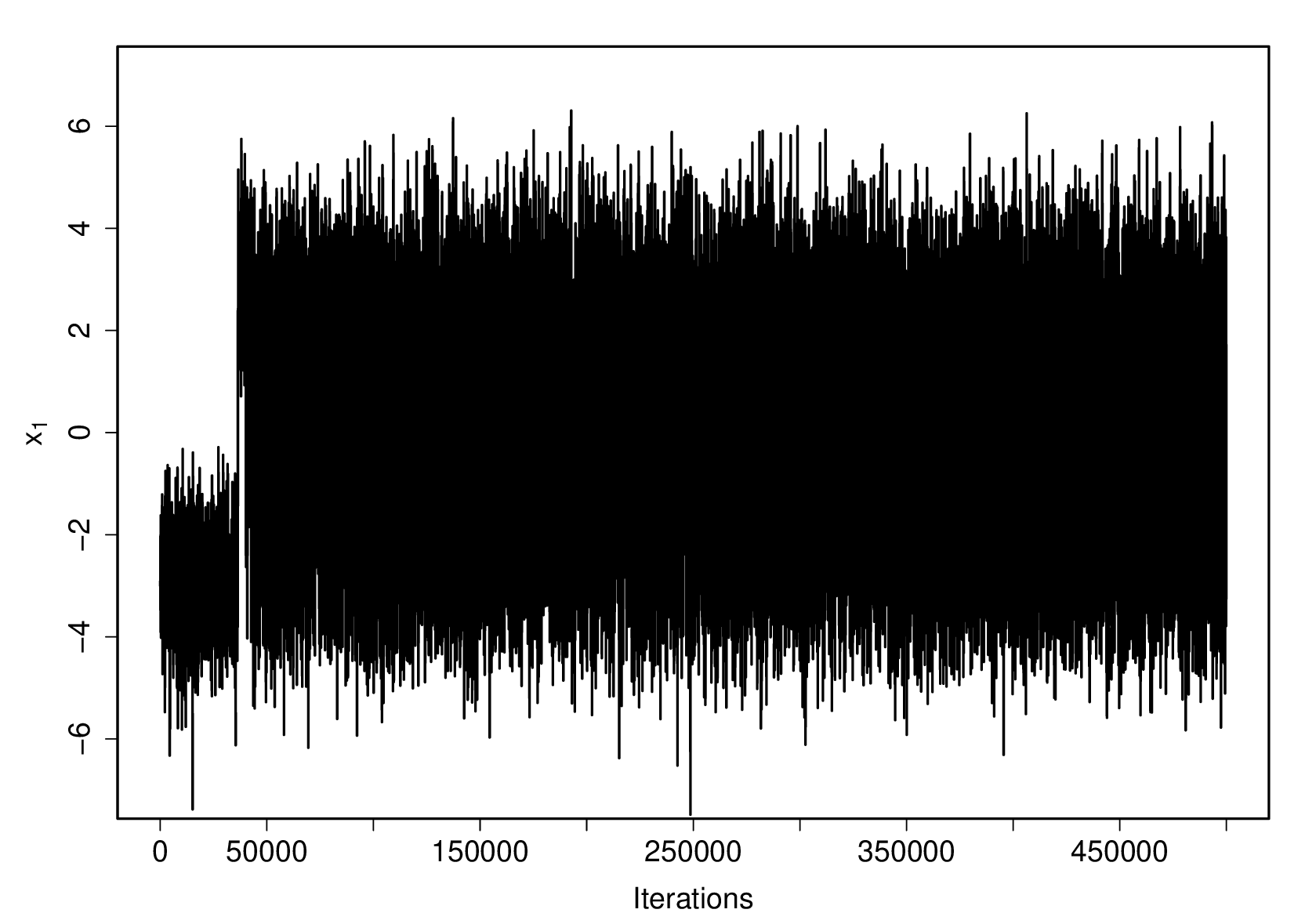}}\\
\subfigure[RWM: wrong marginal]{\includegraphics[scale=0.25,angle=-90]{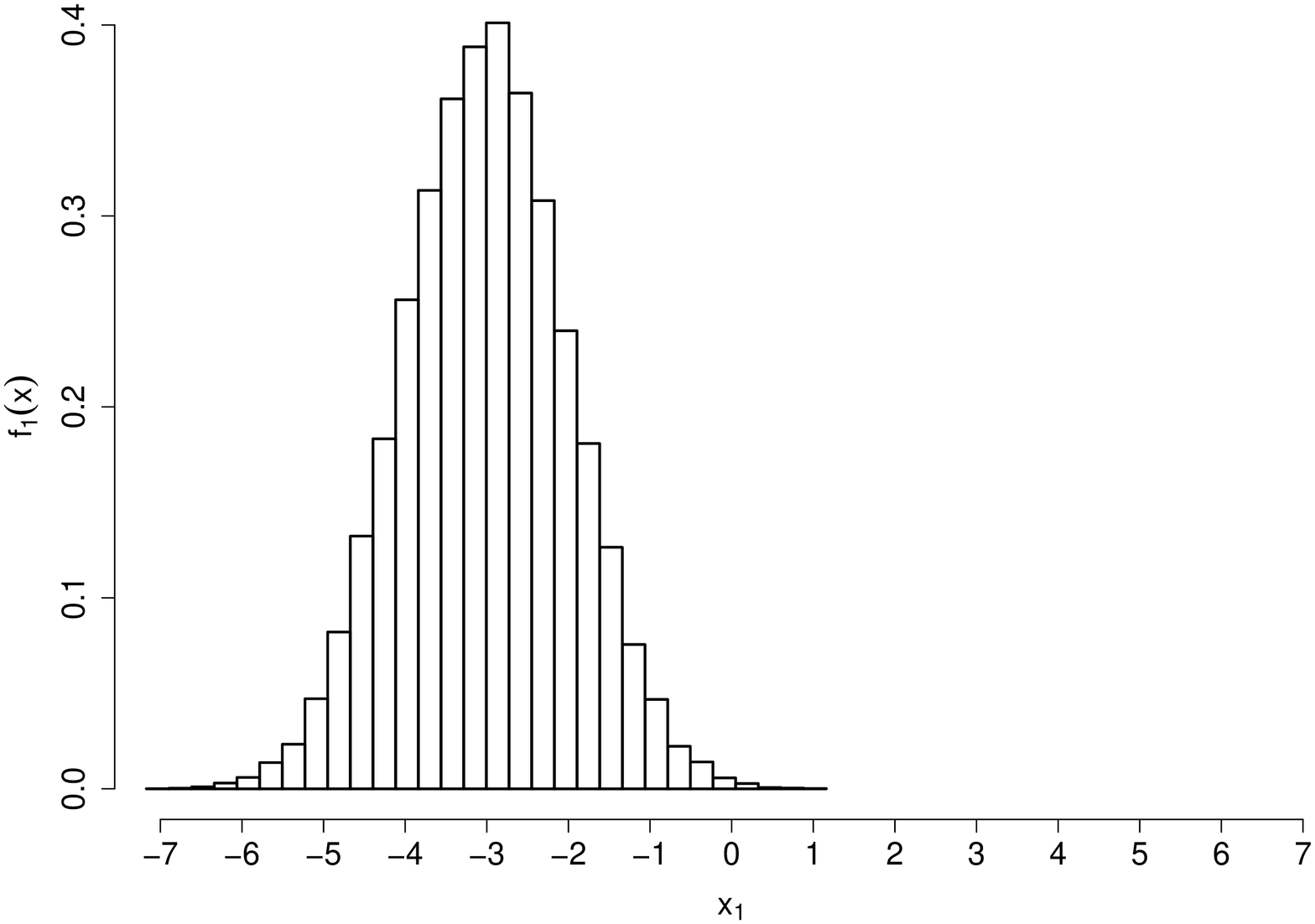}}%
\subfigure[RWM3C: correct marginal]{\includegraphics[scale=0.25,angle=-90]{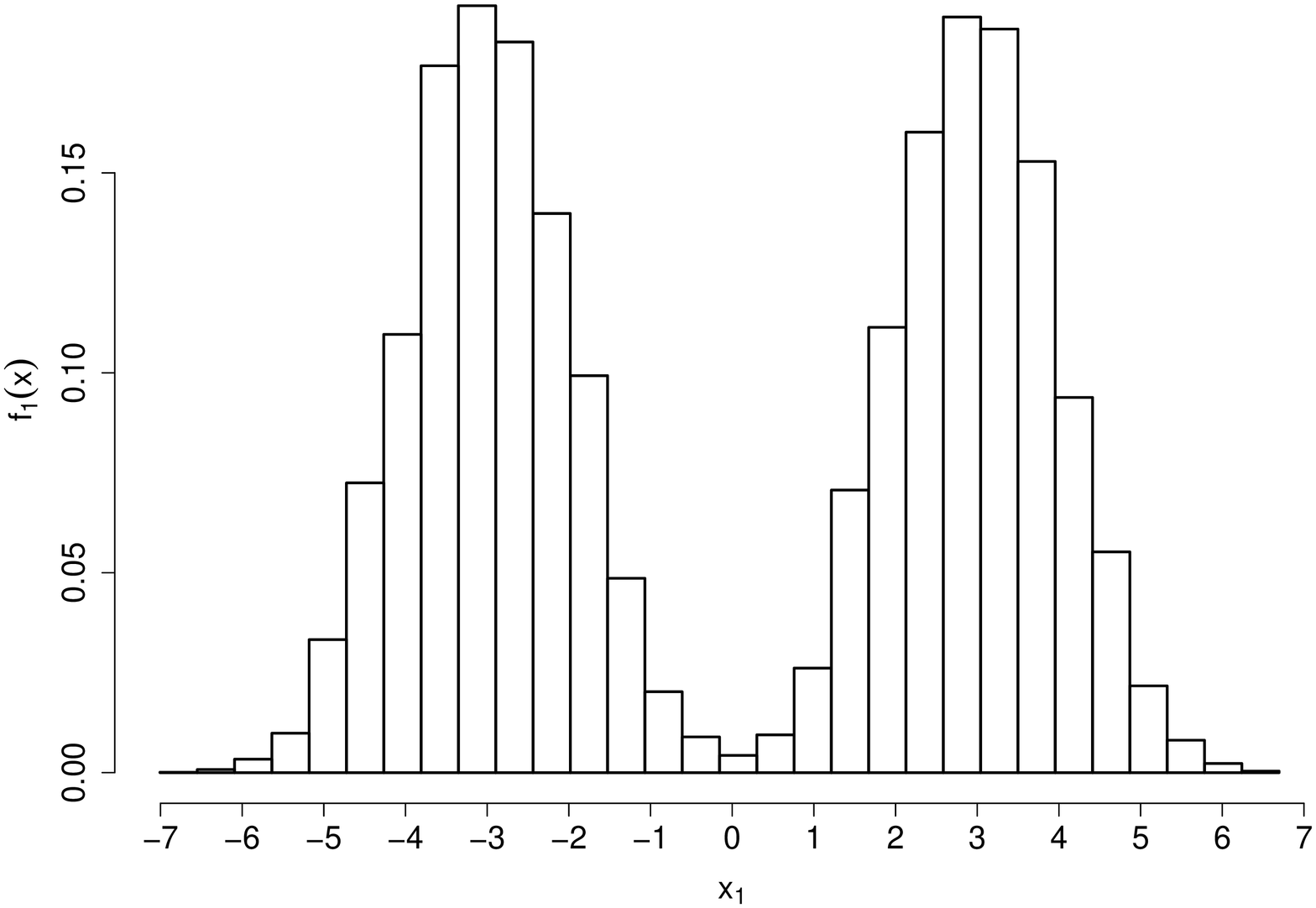}}
\caption{Plot of the iterates and histograms of a sample from a
marginal distribution of a bimodal 5-dimensional mixture of normals distribution using the two component adaptive random walk Metropolis (left panels),
and the three component adaptive random walk Metropolis (right panels).}
\label{fig:bimodal:normal}
\end{figure}

Figure~\ref{fig:bimodal:normal} compares the results and shows that
the two component adaptive random walk fails to explore the
posterior distribution even after 500, 000 iterations, whereas the
three component adaptive random walk can get out of the local modes.

\subsection{A mixture of normals based proposal density}\label{SS: aimh}
The proposal density of the adaptive independent \MH{} approach of %Giordani and Kohn~(2008)
\cite{gior:kohn:2008}
is a mixture with four terms of the form
\begin{align}\label{eq:aimh proposal}
 q_n(x) & = \sum_{j=1}^4\omega_{jn}g_{j}(x| \lambda_{jn} ) \ ,
\quad  \quad \omega_{jn} \ge 0, \quad\text{for} \quad j=1, \dots, 4 \quad \text{and}\quad  \sum_{j=1}^4 \omega_{jn} = 1\ ,
\end{align}
with $\lambda_{jn}$ the  parameter vector for the density $g_{jn}(x;\lambda_{jn})$.
The sampling scheme is run in two stages, which are described below.
Throughout each stage, the parameters in the first two terms are kept fixed.  The first term $g_{1}(x| \lambda_{1n})$ is an estimate of the target
density and the second term $g_{2}(x| \lambda_{2n})$ is a heavy tailed version of $g_{1}(x| \lambda_{1n})$.
The third term $g_{3}(x| \lambda_{3n})$ is an estimate of the target that is updated or adapted as the simulation progresses and the
fourth term $g_{4}(x| \lambda_{4n})$ is a heavy tailed version of the third term.
In the first stage $g_{1n}(x; \lambda_{1n})$ is a Laplace approximation to
the posterior if it is readily available and works well; otherwise, $g_{1n}(x; \lambda_{1n})$ is a Gaussian density constructed from a
preliminary run of 1000 iterates or more of the three component adaptive random
walk.  Throughout, $g_{2}(x| \lambda_{2n})$
has the same component means and probabilities as $g_{1}(x| \lambda_{1n})$, but its
component covariance matrices are ten times those of $g_{1}(x| \lambda_{1n})$. The
term $g_{3}(x| \lambda_{3n})$ is a mixture of normals and $g_{4}(x| \lambda_{4n})$
is also a mixture of normals
obtained by taking its component probabilities and means equal to those of
$g_{3}(x| \lambda_{3n})$, and its component covariance matrices equal to 20 times
those of $g_{3}(x| \lambda_{3n})$. The first stage begins by using $g_{1}(x|
\lambda_{1n})$ and $g_{2}(x| \lambda_{2n})$  only with,  for example, $\omega_{1n}
= 0.8$ and $\omega_{2n} = 0.2$, until there is a
sufficiently large number of iterates to form $g_{3}(x| \lambda_{3n})$.
After that we set $\omega_{1n} = 0.15, \omega_{2n} = 0.05, \omega_{3n} = 0.7 $
and $\omega_{4n} = 0.1$. We begin with a single normal density for $g_{3}(x| \lambda_{3n})$
 and as the simulation progresses we add more components up to a maximum of four according to a
schedule that depends on the ratio of the number of accepted draws to the
dimension of $x$. See~Appendix~\ref{S:sim details}.

In the second stage, $g_{1}(x| \lambda_{1n})$ is set to the value of
$g_{3}(x| \lambda_{3n})$ at the end of the first stage and $g_{2}(x|
\lambda_{2n})$ and $g_{4}(x| \lambda_{4n})$ are constructed as
described above. The heavy-tailed densities $g_{2}(x| \lambda_{2n})$
and $g_{4}(x| \lambda_{4n})$ are included as a defensive strategy,
as suggested by \cite{hest:1995}, to get out of  local modes and to
explore the sample space of the target distribution more
effectively.

It is too computationally expensive to update $g_{3}(x| \lambda_{3n})$ (and hence
$g_{4}(x| \lambda_{4n})$) at every iteration so we
update them according to a schedule that depends on the problem and the
size of the parameter vector. See~Appendix~\ref{S:sim details}.

Our article estimates the multivariate mixture of normals density
for the third component using the method of \cite{gior:kohn:2008}
who identify the marginals that are not symmetric and estimate their
joint density by a mixture of normals using k-harmonic means
clustering. We also estimated the third component using stochastic
approximation but {\em without} first identifying the  marginals
that are not symmetric. We studied the performance of both
approaches to fitting a mixture of normals and found that the
clustering based approach was  more robust in the sense that it does
not require tuning for particular data sets to perform well, whereas
for the more challenging target distributions it was necessary to
tune the parameters of the stochastic approximation to obtain
optimal results. The results for the stochastic approximation
approach are reported in \cite{silva:kohn:giordani:mun2008}. We note
that estimating a mixture of normals using the EM algorithm also
does not require tuning; see \cite{mcla:peel:2000}, as well as
\cite{titterington:1984} for the online EM. However, the EM
algorithm is more sensitive to starting values and is more prone to
converge to degenerate solutions, particularly in an MCMC context
where small clusters of identical observations arise naturally; see
\cite{gior:kohn:2008} for a discussion.

\subsection{A mixture of a $t$ copula and a multivariate $t$ proposal distribution}
\label{SS: TCT}
 The third proposal distribution is a mixture of a
$t$ copula and a multivariate $t$ distribution. We use the $t$
copula as the major component of the mixture because it provides a
flexible and fast method for estimating a multivariate density. In
our applications this means that we assume that after appropriately
transforming each of the parameters, the joint posterior density is
multivariate $t$. Or more accurately, such a proposal density
provides a more accurate estimate of the posterior density than
using a multivariate normal or multivariate $t$ proposal. The second
component in the mixture is a multivariate $t$ distribution with low
degrees of freedom whose purpose is to help the sampler move out of
local modes and  explore the parameter space more effectively.

\cite{joe1997} and %Nelsen~(2006)
\cite{nelsen2006} provide an introduction to copulas,
with details of the $t$ copula given in \cite{demartamcneil2005}.

Let $t_{d,\nu}(\bfz\mid \bfmu ,\bfSigma )$ be a $d$-dimensional
$t$  density  with location $\bfmu$, scale matrix $\bfSigma$
and degrees of freedom $\nu$ and
let  $T_{d,\nu}(\bfz|\bfmu,\bfSigma)$ be the corresponding
cumulative distribution function. Let $f_j(x_j|\bftheta_j)$ and $F_j(x_j|\bftheta_j)$
be the probability density function and the cumulative distribution function
of the $j$th marginal, with $\bftheta_j$ its parameter vector.
Then the density of the $t$  copula based distribution for $\bfx$ is
\begin{align}  \label{eq:student:t:copula}
 g(\bfx) & =\frac{t_{d,\nu}(\bfz;0,\bfSigma_z)}
{\prod_{j=1}^{d}t_{1,\nu}(z_j;
0,1)}\prod_{j=1}^{d}f_j(x_j|\bftheta_j)\ ,
\end{align}
where $z_j$ is determined by $T_{1,\nu}(z_j; 0,1) =
F_j(x_j|\theta_j)$, $j=1, \dots, d$.  For a given sample of
observations, or in our case a sequence of draws, we fit the copula
by first estimating each of the marginals as a mixture of normals.
For the current \dof{}, $\nu_c$, of the $t$ copula, we now transform
each observation $\bfx = (x_1, \dots, x_d)$ to $\bfz = (z_1, \dots,
z_d)$ using
\begin{align} \label{eq:student:t:copula:trans}
T_{1,\nu_c}(z_j;0,1)=F_j(x_j|\bftheta_j)\hspace{0.2cm}\text{for}\hspace{0.2cm}
j=1,\ldots,d.
\end{align}
This produces a sample of $d$ dimensional observations $\bfz$ which we use to
estimate the
scale matrix $\bfSigma_z$ as %$(\nu-2)/\nu$ times
the sample correlation matrix
of the $\bfz$'s. Given the estimates of the marginal distributions and the scale
matrix $\bfSigma_z$, we estimate the degrees of freedom
$\nu$ by maximizing the profile likelihood function given by
\eqref{eq:student:t:copula} over a grid of values $\nu=\{3,5,10,1000\}$,
with $\nu=1000$ representing  the Gaussian copula. We could use instead the grid
of $\nu$ values in \eqref{eq:student:t:copula:trans}, but this is more
expensive computationally and we have found our current procedure works well in practice.

The second component of the mixture is a multivariate $t$ distribution with its
degrees of freedom $\nu_f = 5$ fixed and small,
and with its location and scale parameters estimated from the $x$ iterates using
the first and second sample moments.
The $t$ copula component of the mixture has a weight of 0.7 and the multivariate $t$ component a weight of 0.3.

To draw an observation from the $t$ copula, we first draw $\bfz$
from the multivariate $t$ distribution with location 0 and scale
matrix $\bfSigma_z$ and degrees of freedom $\nu$. We then use a
Newton-Raphson root finding routine to obtain $x_j$ for a given
$z_j$ from \eqref{eq:student:t:copula:trans} for $j=1, \dots, d$.
The details of the computation for the $t$ copula and the schedule
for updating the proposals is given in Appendix~\ref{S:sim details}.

Using antithetic variables in simulations often increases sampling efficiency;
e.g.~\cite{rubinstein:1981}. \cite{tierney1994} proposes using  antithetic variables in \MCMC{} simulation when
the target is symmetric. We apply antithetic variables to the copula based approach which generalizes
Tierney's suggestion by allowing for nonsymmetric marginals.
To the best of our knowledge, this has not been done before.
The antithetic approach is implemented as follows.
As above, we determine probabilistically whether to sample from the copula component or the multivariate $t$ component.
If the copula component is chosen, then $\bfx$ is generated as above and
$\tilde {\bfx} = - \bfx$ is also computed. The values $\bfx$ and $\tilde {\bfx}$
are then transformed to $\bfz$ and $\tilde {\bfz} $ respectively and are accepted or rejected one at a time using the \MH{} method. If we decide to sample from the multivariate $t$ component to obtain $\bfz$, we also compute  $\tilde {\bfz}  = 2 \bfmu - \bfz $, where $\bfmu$ is the mean of the multivariate $t$,
and accept or reject each of these values one at a time using the \MH{} method.

We note that to satisfy the conditions for convergence in
\cite{gior:kohn:2008} we would run the sampling scheme in two
stages, with the first stage as above. In the second stage we would
have a three component proposal, with the first two components the
same as above. The third component would be fixed throughout the
second stage and would be the second component density at the end of
the first stage. The third component would have a small probability,
e.g. 0.05. However, in our examples we have found it unnecessary to
include such a third component as we achieve good performance
without it.

\section{Algorithm comparison}\label{S: algcompar}
This section studies  the  five algorithms  discussed in
Section~\ref{section:adap:samp}. The two component adaptive random
walk metropolis (RWM) and the three component adaptive random walk
Metropolis (RWM3C) are described in Section~\ref{SS: arwm}. The
adaptive independent \MH{} with a mixture of normals proposal
distribution fitted by clustering (IMH-MN-CL) described in
Section~\ref{SS: aimh}. The adaptive independent \MH{}  which is a
mixture of a $t$ copula and a multivariate $t$ proposal, with the
marginal distributions of the copula estimated by mixtures of
normals that are fitted by clustering (IMH-TCT-CL). The fifth sampler
is the antithetic variable version of IMH-TCT-CL, which we call IMH-TCT-CL-A.
These proposals are described in Section~\ref{SS: TCT}.

Our study compares the performance of the algorithms in terms of the acceptance rate of the \MH{} method, the
inefficiency factors (IF) of the parameters, and an overall measure of
effectiveness which compares the
times taken by all samplers to obtain the same level of accuracy.
We define the acceptance rate as the percentage of accepted values of each of the
\MH{} proposals. We define the inefficiency of the sampling scheme for a
given parameter as the variance of the parameter estimate divided by its variance
when the sampling scheme generates independent iterates. We estimate the inefficiency factor as
\begin{equation}
\text{IF} = 1+2\sum_{j=1}^{T}K_{TR}\left(\dfrac{j}{T}\right)\hat{\rho}_j
\label{eq:ineffic}
\end{equation}
where $\hat{\rho}_j$ is the estimated autocorrelation at lag $j$ and the
truncated kernel function $K_{TR}(x) = 1 $
if $|x|\leqslant  1$ and 0 otherwise \citep{andrews1991}.
As a rule of  thumb, the maximum
number of lags $T$ is given by the lowest index $j$ such that
$|\hat{\rho}_j|<2/\sqrt{M}$ with $M$
being the sample size use to compute $\hat{\rho}_j$.
We define the equivalent sample size as $ESS = N/IF$, where $N
=10000$, which can be interpreted as $N$ iterates of the dependent
sampling scheme are equivalent to $ESS$ independent iterates. The
acceptance rate and the inefficiency factor do not take into account
the time taken by a sampler. To obtain an overall measure of the
effectiveness of a sampler, we define its equivalent computing time
$ ECT = N \times IF \times T$, where $T$ is the time per iteration
of the sampler and $N = 100 000$. We interpret $ECT$ as the time
taken by the sampler to attain the same accuracy as that attained by
$N$ independent draws of the same sampler. For two samplers $a$ and
$b$,  $ECT_a/ECT_b$ is the ratio of times taken by them to achieve
the same accuracy.

We note that the time per iteration for a given sampling algorithm
 depends to an important extent on how the
algorithm is implemented, e.g. language used, whether operations are
vectorized, which affects $ECT$ but not the acceptance rates nor
the inefficiencies.

\subsection{Logistic Regression}\label{SS: logistic regression}
This section applies the adaptive sampling schemes to the binary logistic
regression model
\begin{align}\label{e:logistic regression}
 p(y_i = 1|\bfxi,\bfbeta) = \dfrac{\exp\left ( \bfxi'\bfbeta\right )}{1+\exp\left (
 \bfxi'\bfbeta\right ) },
\end{align}
using three different priors for the vector of coefficients $\bfbeta$.
The first is a non-informative multivariate normal prior,
\begin{align} \label{e: normal prior}
\bfbeta & \sim N(0,10^6 I )\ .
\end{align}
The second is a normal prior for the intercept $\beta_0\sim N(0,10^6)$, and a
double exponential, or Laplace prior, for all the other coefficients,
\begin{align}\label{e: dexp}
\beta_{j}\sim \frac{1}{2\tau} \exp \biggl ( - \frac{|\beta_j|}{\tau} \biggr ) \ .
\end{align}
The regression coefficients are assumed to be independent a priori.
We note that this is the prior implicit in the lasso
\citep{tibshirani1996}. The prior for $\tau$ is $\IG(0.01,0.01)$, where $\IG(a,b)$ means
an inverse gamma density with shape $a$ and scale $b$.
The double exponential prior has a spike at zero and heavier tails than the
normal prior. Compared to their posterior distributions under
a diffuse normal prior, this prior shrinks the posterior
distribution of the coefficients close to zero to values even closer
to zero,  while the coefficients far from zero are almost
unmodified. In the adaptive sampling schemes we work with $\log
(\tau) $ rather than $\tau$ as it is unconstrained.

The third prior distribution takes the prior for the intercept as $\beta_0 \sim
N(0,10^6)$ and the prior for the coefficients $\beta_j, j \ge 1,$ as the
two component mixture of normals,
\begin{align} \label{e: MN prior}
\beta_j & \sim \omega \phi_1(\beta_j; 0, \tau_S^2) + (1-\omega) \phi_1(\beta_j; 0,
\tau_L^2)\ ,
\end{align}
with the regression coefficients assumed a priori independent.
\cite{george_mcculloch1993} suggest using this prior for Bayesian
variable selection, with $\tau_S^2$ and  $\tau_L^2$ small and large
variances that are chosen by the user. In our article their values
are given for each of the examples below. The prior for $\omega$ is
uniform. In the adaptive sampling we work with the logit of $\omega$
because it is unconstrained.

%\begin{comment}
% MROZ Data - Logistic Regression
%------------------------------------------------
\noindent
\section*{Labor force participation by women}
This section models the probability of labor force participation by
women, $inlf$, in 1975 as a function of the covariates listed in
Table~\ref{t: labor force part}. This data set is
discussed by %Wooldridge~~(2007, p. 537)
\cite{wool:2004}, p.~537 and has a sample size of 753.
\begin{table}[!ht]
\centering
\caption{Variables used in labor force participation data regression}
 \begin{tabular}{l l}\hline
 $inlf:$          & 1 if the woman is in labor force in 1975 and 0 otherwise; \\
 $kidslt6:$       & 1 if kids are with less than 6 years and 0 otherwise;\\
 $kidsge6:$       & 1 if kids are between 6 and 18 years and 0 otherwise;\\
 $age:$           & Age of the woman;\\
 $educ:$          & Years of schooling;\\
 $hushrs:$        & Hours worked by husband in 1975;\\
 $huswage:$       & Husband's hourly wage in 1975;\\
 $mtr:$           & Federal marginal tax rate facing woman;\\
 $exper:$         & Actual labor market experience; and\\
 $nwifeinc:$      & family income ($faminc$) in 1975 minus wage times hour divided by 1000\\
                  & ($(faminc - wage\times hours)/1000$).\\\hline
 \end{tabular}
\label{t: labor force part}
 \end{table}

We ran all the adaptive sampling schemes
presented in Section \ref{section:adap:samp} for all three prior distributions.
Our targets are 12 dimensional (normal diffuse prior)  and 13
dimensional (double exponential prior and mixture of normals prior) posterior distributions.
Our starting values and initial proposal distributions
for all the adaptive algorithms were obtained by fitting a
generalized linear model using the function \textsf{glmfit} in \MATLAB{}.
We used $\tau_S^2=0.01$ and $\tau_L^2=10000$ for the mixture of normals prior. Table~\ref{table:stats:MROZ} summarizes the posterior distributions
for the three priors.

\begin{table}[!ht]
 \centering
 \caption{Summary statistics for the logistic regression applied to the labor force participation data  under normal, double exponential (with $\theta=\log(\tau)$) and mixture of normals (with $\theta=\text{logit}(\omega)$)
 prior distributions.}
 \begin{tabular}{l | cc|cc|cc}\hline\hline
 Parameter         & \multicolumn{2}{c|}{Normal}&\multicolumn{2}{c|}{Double Exponential}&
 \multicolumn{2}{c}{Mixture of Normals}\\\cline{2-7}
                   & Mean     & S. Dev.&  Mean   & S. Dev.& Mean    & S. Dev.\\\hline
 $\theta$          &    -     &  -     &  0.6107 & 0.3973 &  1.3208 & 0.7118 \\
 intercept         &  22.4612 & 3.1836 & 16.5521 & 3.7459 & 23.6789 & 2.8201 \\
 $kidslt6$         &  -1.0685 & 0.2200 & -1.0885 & 0.2157 & -1.1618 & 0.2168 \\
 $kidsge6$         &   0.3347 & 0.0862 &  0.2864 & 0.0884 &  0.1763 & 0.0649 \\
 $age$             &  -0.0688 & 0.0164 & -0.0703 & 0.0158 & -0.0790 & 0.0151 \\
 $educ$            &   0.1521 & 0.0492 &  0.1612 & 0.0480 &  0.1225 & 0.0424 \\
 $hushrs$          &  -0.0010 & 0.0002 & -0.0009 & 0.0002 & -0.0008 & 0.0002 \\
 $huswage$         &  -0.2587 & 0.0522 & -0.2220 & 0.0530 & -0.1851 & 0.0436 \\
 $mtr$             & -23.2281 & 3.5870 &-16.2031 & 4.3158 &-25.1363 & 3.0601 \\
 $exper$           &   0.7621 & 0.1584 &  0.8448 & 0.1648 &  0.1944 & 0.0584 \\
 $nwifeinc$        &  -0.1355 & 0.0241 & -0.1074 & 0.0254 & -0.1211 & 0.0212 \\
 $exper^2$         &  -0.0030 & 0.0012 & -0.0029 & 0.0012 & -0.0026 & 0.0011 \\
 $mtr\times exper$ &  -0.8276 & 0.2219 & -0.9472 & 0.2312 & -0.0419 & 0.0781 \\
\hline\hline
\end{tabular}
\label{table:stats:MROZ}
\end{table}

Table~\ref{t: mroz effic truncated}  shows the acceptance rates,
inefficiencies, the equivalent sample sizes and the equivalent
computing times  of the adaptive sampling schemes for the three
prior distributions.

 \begin{table}[!ht]
 \centering
 \caption{Acceptance rates, inefficiencies (IF) based on the truncated kernel,
 equivalent sample size (ESS=10000/IF) and ECT = IF $\times$ time for 100 000 iterations
 for the logistic regression model applied to the labor force participation data.}
 {\footnotesize
 \begin{tabular}{l | c | rrr | rrr | rrr}\hline\hline
 & & \multicolumn{3}{c|}{Inefficiency}  & \multicolumn{3}{c|}{Equivalent sample size}
 &\multicolumn{3}{c}{ECT}   \\\cline{3-11}
 Algorithm  & A. Rate  & Min & Median & Max & Min & Median & Max & Min & Median & Max \\\hline
 & \multicolumn{10}{c}{Normal Prior}    \\\hline
 RWM             & 16.9 & 63.149 & 66.419 & 69.542 &  144 &   151 &   158 & 1696 & 1784 & 1868 \\
 RWM3C           & 30.4 & 45.098 & 46.744 & 48.685 &  205 &   214 &   222 & 1219 & 1264 & 1316 \\
 IMH-MN-CL       & 67.1 &  1.981 &  2.188 &  2.426 & 4121 &  4571 &  5049 &  191 &  211 &  234 \\
 IMH-TCT-CL      & 76.5 &  1.695 &  1.761 &  1.918 & 5213 &  5680 &  5900 &  161 &  167 &  182 \\
 IMH-TCT-CL-A    & 79.2 &  0.749 &  0.836 &  0.938 & 10655 &  11959 &  13349 &  202 &  226 &  253 \\
  \hline
 & \multicolumn{10}{c}{Double exponential Prior}    \\\hline
 RWM             & 16.7 & 68.437 & 71.198 & 75.024 &  133 &   140 &   146 & 1907 & 1984 & 2090 \\
 RWM3C           & 30.1 & 47.749 & 50.993 & 52.587 &  190 &   196 &   209 & 1340 & 1431 & 1476 \\
 IMH-MN-CL       & 66.6 &  2.080 &  2.185 &  2.415 & 4140 &  4578 &  4807 &  258 &  271 &  299 \\
 IMH-TCT-CL      & 76.2 &  1.753 &  1.874 &  2.614 & 3826 &  5336 &  5706 &  277 &  296 &  413 \\
 IMH-TCT-CL-A    & 77.7 &  1.007 &  1.182 &  1.896 & 5273 &  8460 &  9935 &  263 &  308 &  495 \\
 \hline
           & \multicolumn{10}{c}{Mixture of Normals Prior}    \\\hline
 RWM             & 25.9 & 36.243 & 44.711 & 46.234 &  216 &   224 &   276 & 1230 & 1517 & 1568 \\
 RWM3C           & 29.7 & 49.411 & 52.163 & 79.618 &  126 &   192 &   202 & 1681 & 1774 & 2708 \\
 IMH-MN-CL       & 68.6 &  1.986 &  2.126 &  2.642 & 3786 &  4703 &  5036 &  204 &  218 &  271 \\
 IMH-TCT-CL      & 76.7 &  1.643 &  1.776 &  2.030 & 4926 &  5630 &  6085 &  352 &  381 &  435 \\
 IMH-TCT-CL-A    & 81.8 &  0.661 &  0.733 &  1.136 & 8800 &  13651 &  15120 &  164 &  182 &  282 \\
 \hline\hline
 \end{tabular}
 }
 \label{t: mroz effic truncated}
 \end{table}

 Although the acceptance rates for the three component adaptive
random walk are higher than for the two component adaptive random
walk, the results are more ambiguous when comparing their
inefficiencies. The inefficiencies for all the adaptive independent
\MH{} schemes are much lower than the two adaptive
random walk Metropolis schemes.
The copula based proposal distribution
performed the best in terms of the acceptance rates and the
inefficiency factors, and its antithetic version performed even better.
%\end{comment}
%------------------------------------------------
% HMDA Data - Logistic Regression
%------------------------------------------------
\section*{Home mortgage disclosure act}
The home mortgage disclosure act (HMDA) data set relates to mortgage applications
in Boston in 1990. It is discussed in Section~11.4 of
\cite{stock_watson2007} and has been analyzed by many authors, e.g.
%Megbolugbe and Carr~(1993)
\cite{megb:carr:1993}. The dependent variable is {\it deny}, which is 1 if a
mortgage application is denied and 0 otherwise. The sample size is 2380
with the  covariates listed in Table~\ref{t: HMDA data}.

\begin{figure}[!ht]
\centering \subfigure[Marginal posterior distribution for
$\beta_{black}$]
{\includegraphics[scale=0.27,angle=-90]{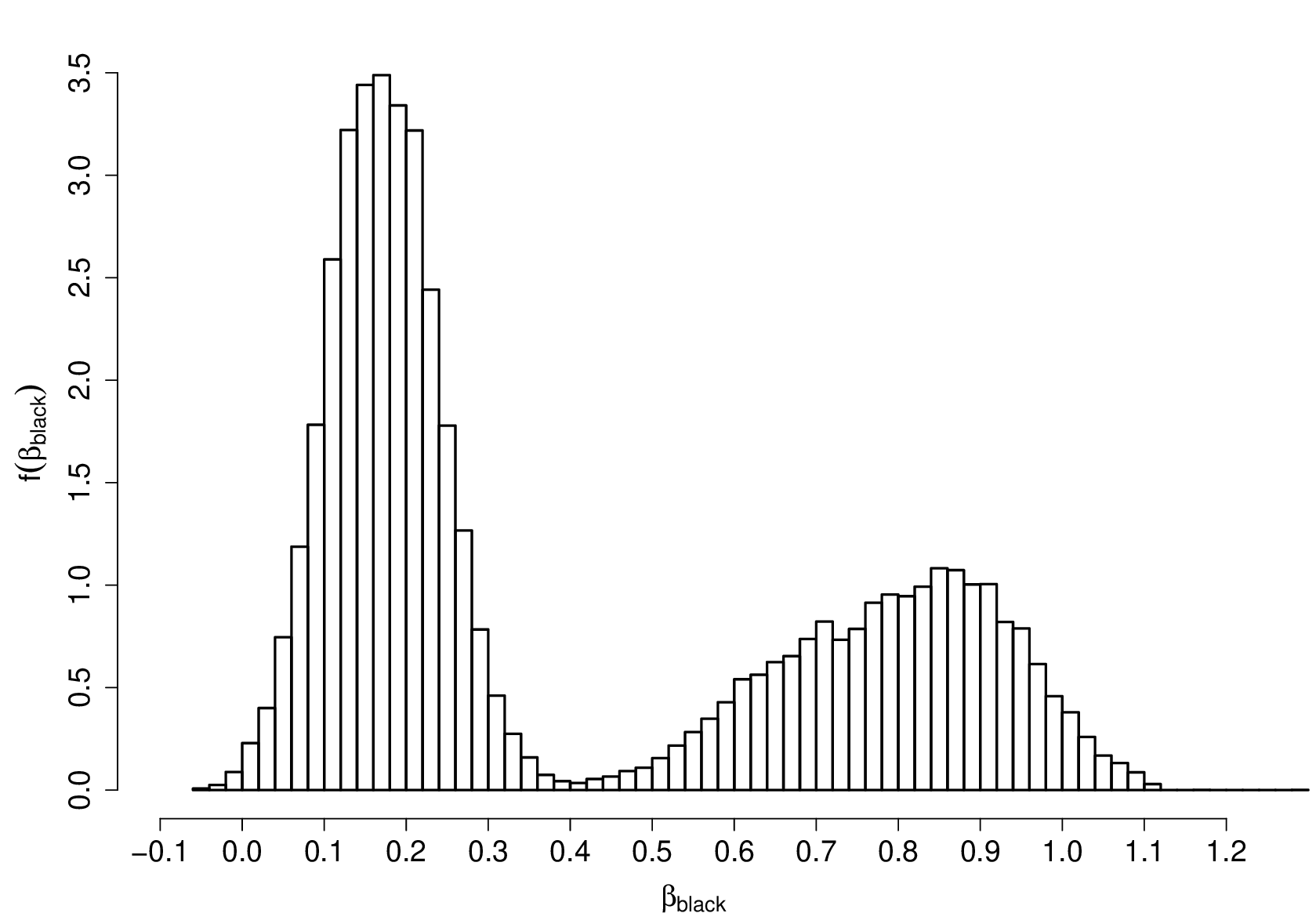}}
\subfigure[Marginal posterior distribution for $\beta_{ltvhigh}$]
{\includegraphics[scale=0.27,angle=-90]{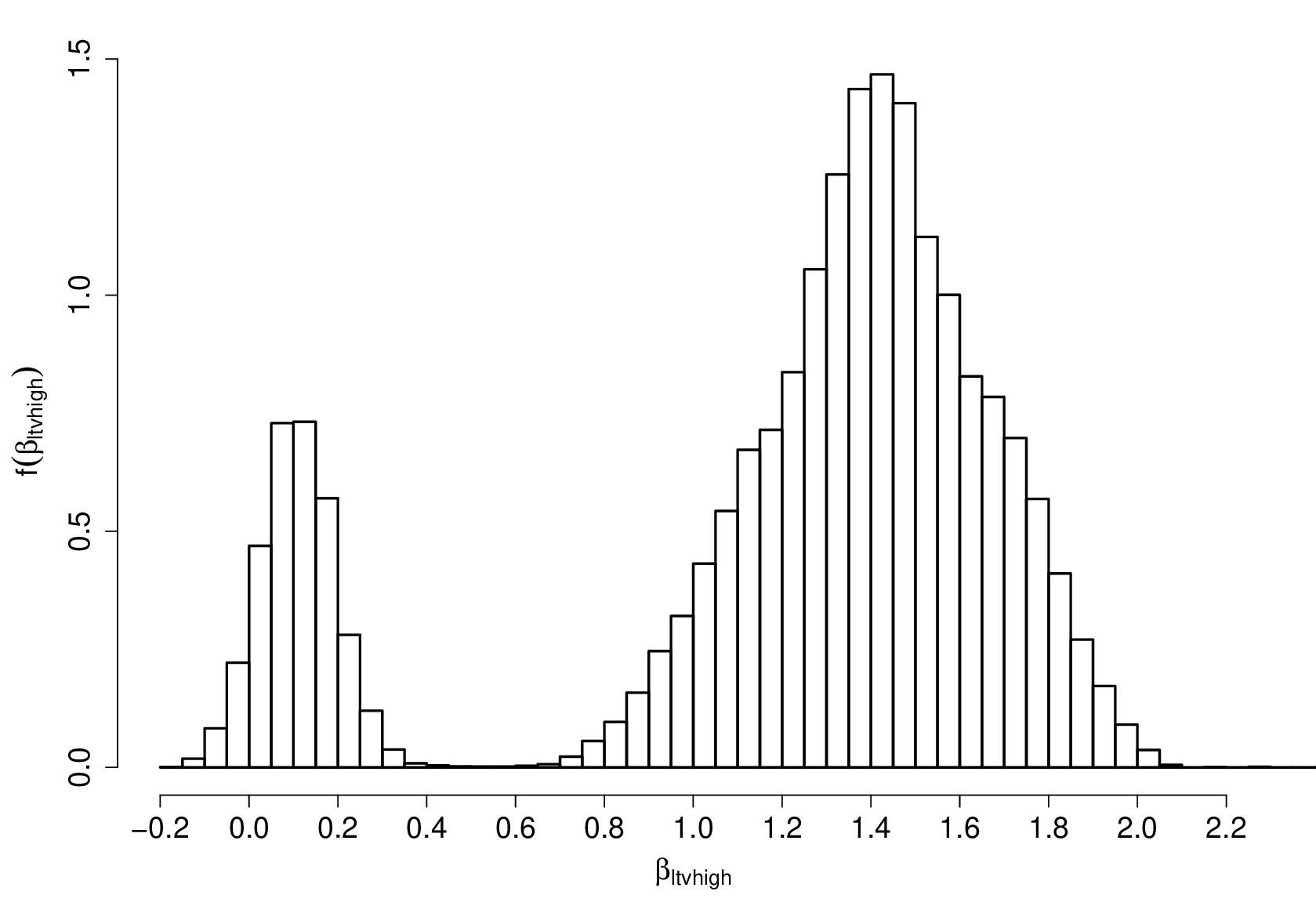}}
\caption{Two marginal posterior distributions of the logistic
regression model applied to the home mortgage disclosure act data
under mixture of normal prior with $\tau_S^2=0.01$ and $\tau_L^2=10000$.}
\label{fig:bimodal:hmda}
\end{figure}

 \begin{table*}[!ht]
 \centering
 \caption{Covariates used in the HMDA regression}
 \begin{tabular}{l l}\hline
 $pirat:$   & Scaled total obligation to income ratio;\\
 $black:$   & 1 if African-American and 0 otherwise;\\
 $ccred:$   & Credit history of consumer payments:\\
            & \hspace{1cm}1 if no slow pay,\\
            & \hspace{1cm}2 if one or two  slow pay accounts,\\
            & \hspace{1cm}3 if more than two slow pay accounts, and\\
            & \hspace{1cm}4 if insufficient credit history;\\
 $mcred:$   & Mortgage history:\\
            & \hspace{1cm}1 if no late mortgage payments,\\
            & \hspace{1cm}2 if no mortgage payment history,\\
            & \hspace{1cm}3 if one or more late mortgage payments, and\\
            & \hspace{1cm}4 if More than two late mortgage payments;\\
 $pubrec:$  & 1 if there is a bankruptcy public record and 0 otherwise;\\
 $denpmi:$  & 1 if a private mortgage insurance has been denied and 0 otherwise;\\
 $selfemp:$ & 1 if self employed and 0 otherwise;\\
 $married:$ & 1 if married and 0 otherwise;\\
 $hischl:$  & 1 if number of years of schooling $\geqslant$ 12 and 0 otherwise;\\
 $lvtmed:$  & (loan-to-value $\geqslant$ 80\%) $\times$ (loan-to-value $\leqslant$ 95\%);\\
 $ltvhigh:$ & loan-to-value $>$ 95\%; \\
 $ccred3:$  & 1 if credit history of consumer payments is equal to 3 and 0 otherwise;\\
 $ccred4:$  & 1 if credit history of consumer payments is equal to 4 and 0 otherwise;\\
 $ccred5:$  & 1 if credit history of consumer payments is equal to 5 and 0 otherwise; and\\
 $ccred6:$  & 1 if credit history of consumer payments is equal to 6 and 0 otherwise.\\\hline
\end{tabular}
\label{t: HMDA data}
\end{table*}
Our starting values and initial proposal distributions
for all the adaptive algorithms were obtained by fitting a
generalized linear model using the function \textsf{glmfit} in \MATLAB{}.
We used $\tau_S^2=0.01$ and $\tau_L^2=10000$ for the mixture of normals
prior. Table~\ref{table:stats:HMDA} summarizes the estimation
results and Figure~\ref{fig:bimodal:hmda} shows the two marginals  that are
bimodal under the mixture of normals prior.

 \begin{table}[!ht]
 \centering
 \caption{Summary statistics for the logistic regression applied to the home mortgage
 disclosure act data under normal, double exponential (with $\theta=\log(\tau)$) and mixture of normals (with $\theta=\text{logit}(\omega)$)  prior distributions. }
 \begin{tabular}{l | cc|cc|cc}\hline\hline
 Parameter         & \multicolumn{2}{c|}{Normals}&\multicolumn{2}{c|}{Double Exponential}&
 \multicolumn{2}{c}{Mixture of Normals}\\\cline{2-7}
            & Mean    & S. Dev.&  Mean   & S. Dev.& Mean    & S. Dev.\\\hline
 $\theta$   &    -    &  -     &  0.1637 & 0.2890 & 0.8536 & 0.5573 \\
 intercept  & -4.9153 & 0.6744 & -4.4802 & 0.6342 &-5.0159 & 0.3489 \\
 $pirat$    &  4.8068 & 0.7904 &  4.2074 & 0.7831 & 4.8814 & 0.7589 \\
 $black$    &  0.6042 & 0.1797 &  0.5955 & 0.1766 & 0.4081 & 0.3190 \\
 $ccred$    &  0.7326 & 0.2134 &  0.4498 & 0.1464 & 0.2718 & 0.0403 \\
 $mcred$    &  0.2215 & 0.1456 &  0.2125 & 0.1428 & 0.1079 & 0.0797 \\
 $pubrec$   &  1.2814 & 0.2132 &  1.2360 & 0.2118 & 1.3572 & 0.1976 \\
 $denpmi$   &  4.7761 & 0.5888 &  4.4736 & 0.5510 & 4.6734 & 0.5538 \\
 $selfemp$  &  0.6645 & 0.2160 &  0.6388 & 0.2143 & 0.1012 & 0.0912 \\
 $married$  & -0.3971 & 0.1544 & -0.3799 & 0.1520 &-0.1290 & 0.0831 \\
 $hischl$   & -1.1721 & 0.4276 & -0.9786 & 0.4315 &-0.0507 & 0.0968 \\
 $lvtmed$   &  0.4933 & 0.1616 &  0.4686 & 0.1614 & 0.1451 & 0.0906 \\
 $ltvhigh$  &  1.5686 & 0.3193 &  1.4515 & 0.3235 & 1.2000 & 0.5342 \\
 $ccred3$   & -0.6192 & 0.4683 & -0.1229 & 0.3454 & 0.0157 & 0.0917 \\
 $ccred4$   & -0.6872 & 0.6469 &  0.0544 & 0.4433 & 0.0556 & 0.0954 \\
 $ccred5$   & -1.7431 & 0.8103 & -0.6463 & 0.5597 &-0.0086 & 0.0906 \\
 $ccred6$   & -2.1088 & 1.0084 & -0.7534 & 0.6892 & 0.0003 & 0.0923 \\
 \hline\hline
 \end{tabular}
 \label{table:stats:HMDA}
 \end{table}

Table~ \ref{t: hmda truncated} shows the acceptance rates,
inefficiencies, the equivalent sample sizes and the \ect{}  of the
adaptive sampling schemes for the three prior distributions. Both
adaptive random walk Metropolis algorithms have much higher
inefficiencies than the adaptive independent \MH{} algorithms,
especially under the mixture of normals prior distribution.  The copula based proposal has the highest acceptance rates and
lowest inefficiencies,
especially for the mixture of normals prior. The reason for this may
be that this prior produces a multimodal posterior distribution.

\begin{table}[!ht]
 \centering
 \caption{Acceptance rates, inefficiencies (IF) based on the truncated kernel,
 equivalent sample size (ESS=10000/IF) and ECT = IF $\times$ time for 100 000 iterations
 for the logistic regression model applied to the home mortgage disclosure act data.}
 {\footnotesize
 \begin{tabular}{l | c | rrr | rrr | rrr}\hline\hline
 & & \multicolumn{3}{c|}{Inefficiency}  & \multicolumn{3}{c|}{Equivalent sample size}
 &\multicolumn{3}{c}{ECT}   \\\cline{3-11}
 Algorithm & A. Rate & Min & Median & Max & Min & Median & Max & Min & Median & Max \\\hline
 & \multicolumn{10}{c}{Normal Prior}    \\\hline
 RWM          & 28.5 & 49.974 & 52.714 & 54.216 &  184 &  190 &  200 & 2872 & 3030 & 3116 \\
 RWM3C        & 30.0 & 59.084 & 61.101 & 63.796 &  157 &  164 &  169 & 3405 & 3521 & 3676 \\
 IMH-MN-CL    & 69.2 &  1.974 &  2.040 &  2.211 & 4522 & 4903 & 5067 &  302 &  312 &  338 \\
 IMH-TCT-CL   & 76.8 &  1.612 &  1.821 &  2.136 & 4682 & 5490 & 6205 &  332 &  375 &  439 \\
 IMH-TCT-CL-A & 63.9 &  1.762 &  2.503 &  4.095 & 2442 & 3995 & 5676 &  519 &  737 & 1206 \\
 \hline       & \multicolumn{10}{c}{Double exponential Prior}    \\\hline
 RWM          & 27.5 & 53.770 & 57.625 & 60.116 &  166 &  174 &  186 & 3153 & 3379 & 3525 \\
 RWM3C        & 29.5 & 61.348 & 63.907 & 68.001 &  147 &  156 &  163 & 3609 & 3760 & 4001 \\
 IMH-MN-CL    & 63.8 &  2.244 &  2.401 &  2.968 & 3369 & 4164 & 4457 &  195 &  208 &  257 \\
 IMH-TCT-CL   & 65.0 &  2.266 &  2.468 &  2.743 & 3645 & 4052 & 4412 &  712 &  775 &  862 \\
 IMH-TCT-CL-A & 65.1 &  1.665 &  1.865 &  2.626 & 3807 & 5361 & 6006 &  627 &  702 &  989 \\
 \hline       & \multicolumn{10}{c}{Mixture of Normals Prior}    \\\hline
 RWM          & 15.8 & 66.962 & 96.965 & 159.411 &  63 &  103 &  149 & 4309 &  6240 &  10258 \\
 RWM3C        & 20.1 & 91.179 &104.422 & 159.808 &  63 &   96 &  110 & 5886 &  6741 &  10316 \\
 IMH-MN-CL    & 22.4 & 15.066 & 26.898 & 109.568 &  91 &  372 &  664 & 1108 &  1978 &   8057 \\
 IMH-TCT-CL   & 59.0 &  3.041 &  3.293 &  13.016 & 768 & 3037 & 3288 & 1520 &  1647 &   6508 \\
 IMH-TCT-CL-A & 55.7 &  3.096 &  3.476 &  13.193 & 758 & 2877 & 3230 & 1934 &  2172 &   8241 \\
 \hline\hline
 \end{tabular}
 }
 \label{t: hmda truncated}
 \end{table}
We remark that we monitored the updates of the
proposal distributions through all iterates of all algorithms using
the $t$ copula  based sampler and noted that most of the time the
degrees of freedom was generated as 1 000, effectively giving a
Gaussian copula.

% Quantile Regression
%--------------------------------------------------------------------------------------------------
\subsection{Bayesian Quantile Regression} \label{SS: quantile regression}
In classical quantile regression the vector of $\beta$ coefficients is estimated
by minimizing the function
\begin{align*}
H_\delta(\bfy|\bfbeta) & = \sum_{i=1}^n
\rho_\delta(y_i-\bfxi'\bfbeta) \ ,
\end{align*}
for $\beta$ for a given $0<\delta<1$, where
$ \rho_\delta(u) = (|u|+(2\delta-1)u)/2$.
We can turn this into a likelihood function based on the asymmetric Laplace distribution, i.e.
\begin{align}\label{e: likelihood quantile regression}
p_\delta(\bfy|\bfbeta,\sigma,\bfx) & = \prod_{i=1}^n
\delta(1-\delta)\sigma^{-1}\exp\left\{-\rho_\delta\left(
\dfrac{y_i-\bfxi^T\bfbeta}{\sigma}\right)\right\} .
\end{align}
See for example \cite{yu:moye:2001} and references therein for a
discussion of Bayesian quantile regression. In the example below, we use the same three
priors  for the regression parameters as in Section~\ref{SS:
logistic regression}, an $\IG(0.01,0.01)$ prior for $\sigma$ and
take  $\tau_S^2=0.0001$ and $\tau_L^2=1$ for the mixture of normals
prior distribution.

%\subsubsection
\section*{Current population survey}  \label{section:CPS}
We consider a data set of full-time  male workers given by the U.S.
current population survey of March 2000. The dependent variable is the logarithm of wage
and salary income and the covariates are listed in Table~\ref{t: CPS}.
The sample size is 5149 observations. Similar data are analyzed by \cite{donald2000}. We carry out Bayesian quantile regression for this data set for $\delta = 0.1$. Similar analyses for $\delta = 0.5$ and $\delta =
0.9$ are reported in \cite{silva:kohn:giordani:mun2008}.
The target distributions in this example are 24 and 25 dimensional. Our initial proposal distributions
for the adaptive independent samplers were obtained by first running the 3 component adaptive random walk
for 20 000 iterations for the normal and double exponential priors and 100 000 iterations for the mixture of normals prior. In all cases we initialized the random walks using linear regression estimates.
Table \ref{table:stats:CPS:q010} summarizes the marginal posterior
distributions of the parameter estimates.

 \begin{table}[!bht]
 \centering
 \caption{Covariates for the quantile regressions of log income in CPS data }
 \begin{tabular}{l l}\hline
 \hspace{0.5cm}
 $age$      & Age of worker in years;\\
  \hspace{0.5cm}
 $age^2$ & Square of age \\
 Education: & \\
 \hspace{0.5cm}
 $edhsinc$  & 1 if incomplete high school and 0 otherwise;\\
 \hspace{0.5cm}
 $edpssome$ & 1 if some post-secondary and 0 otherwise;\\
 \hspace{0.5cm}
 $eddegree$ & 1 if completed college or university and 0 otherwise;\\
            &(Omitted category is completed high school.)\\
 Coverage:  & \\
 \hspace{0.5cm}
 $duncov$   & 1 if covered by a union negotiated collective agreement and 0 otherwise;\\
 Employment:& \\
 \hspace{0.5cm}
 $dpub$     & 1 if works in public sector and 0 otherwise;\\
 Occupation:& \\
 \hspace{0.5cm}
 $occwhtp$  & white collar professional (e.g. managers)  \\
 \hspace{0.5cm}
 $occwhto$  & while collar other (e.g. clerical)\\
            & (Omitted category is blue collar.)\\
 Industry:  & \\
 \hspace{0.5cm}
 $indprim$  & 1 if primary and 0 otherwise;\\
 \hspace{0.5cm}
 $indtrans$ & 1 if transport and commn and 0 otherwise;\\
 \hspace{0.5cm}
 $indtrade$ & 1 if trade and 0 otherwise;\\
 \hspace{0.5cm}
 $indserv$  & 1 if services and 0 otherwise;\\
 \hspace{0.5cm}
 $indpuba$  & 1 if public admin and 0 otherwise;\\
            & (Omitted category is manufacturing sector.)\\
 Location:  & \\
 \hspace{0.5cm}
 $smsam$    & 1 if location in metropolitan area and 0 otherwise;\\
 Region:    & \\
 \hspace{0.5cm}
 $regmw$    & 1 if mid-west and 0 otherwise;\\
 \hspace{0.5cm}
 $regs$     & 1 if south and 0 otherwise;\\
 \hspace{0.5cm}
 $regw$     & 1 if west and 0 otherwise;\\
            & (Omitted category is north east.)\\
 Marital:   & \\
 \hspace{0.5cm}
 $marnev$   & 1 if never married and 0 otherwise;\\
 \hspace{0.5cm}
 $marsdw$   & 1 if separated, divorced or widowed and 0 otherwise;\\
            & (Omitted category is married.)\\
 Dependents:& \\
 \hspace{0.5cm}
 $frelu6$   & 1 if kids are with less than 6 years and 0 otherwise;\\
 \hspace{0.5cm}
 $frelu18$  &  1 if kids are between 6 and 18 years and 0 otherwise; and\\
 Race:      & \\
 \hspace{0.5cm}
 $vm$       & 1 if visible minority and 0 otherwise.\\\hline
 \end{tabular}
 \label{t: CPS}
 \end{table}

\begin{table}[!t]
 \centering
 \caption{Summary statistics for the 0.1-quantile regression applied to the
 current  population survey data under normal, double exponential (with $\theta=\log(\tau)$) and mixture of normals (with $\theta=\text{logit}(\omega)$)  prior distributions. }
 \begin{tabular}{l | cc|cc|cc}\hline\hline
 Parameter         & \multicolumn{2}{c|}{Normals}&\multicolumn{2}{c|}{Double Exponential}&
 \multicolumn{2}{c}{Mixture of Normals}\\\cline{2-7}
            &  Mean    & S. Dev. &  Mean    & S. Dev. &  Mean    & S. Dev. \\\hline
 $\theta$   &     -    &   -     & -5.43902 & 0.21692 &  3.70487 & 1.28372 \\
 $\sigma$   &  0.01032 & 0.00014 &  0.01032 & 0.00014 &  0.01032 & 0.00014 \\
 intercept  &  0.02686 & 0.00825 &  0.03108 & 0.00806 &  0.03299 & 0.00745 \\
 $age$      &  0.02286 & 0.00420 &  0.02056 & 0.00412 &  0.01956 & 0.00377 \\
 $age^2$    & -0.00205 & 0.00052 & -0.00176 & 0.00051 & -0.00164 & 0.00047 \\
 $edhsinc$  & -0.01448 & 0.00196 & -0.01437 & 0.00188 & -0.01435 & 0.00186 \\
 $edpssome$ &  0.00599 & 0.00161 &  0.00571 & 0.00158 &  0.00582 & 0.00155 \\
 $eddegree$ &  0.01696 & 0.00191 &  0.01657 & 0.00190 &  0.01675 & 0.00185 \\
 $duncov$   &  0.01478 & 0.00199 &  0.01420 & 0.00207 &  0.01424 & 0.00197 \\
 $dpub$     & -0.00301 & 0.00291 & -0.00168 & 0.00259 & -0.00208 & 0.00270 \\
 $occwhtp$  &  0.02694 & 0.00204 &  0.02640 & 0.00198 &  0.02586 & 0.00196 \\
 $occwhto$  &  0.00268 & 0.00166 &  0.00215 & 0.00158 &  0.00200 & 0.00158 \\
 $indprim$  & -0.00835 & 0.00206 & -0.00764 & 0.00199 & -0.00767 & 0.00198 \\
 $indtrans$ & -0.00216 & 0.00229 & -0.00138 & 0.00211 & -0.00135 & 0.00222 \\
 $indtrade$ & -0.01902 & 0.00205 & -0.01803 & 0.00198 & -0.01780 & 0.00195 \\
 $indserv$  & -0.01753 & 0.00199 & -0.01677 & 0.00192 & -0.01647 & 0.00188 \\
 $indpuba$  &  0.00711 & 0.00420 &  0.00631 & 0.00385 &  0.00715 & 0.00382 \\
 $smsam$    &  0.00767 & 0.00134 &  0.00734 & 0.00134 &  0.00748 & 0.00132 \\
 $regmw$    &  0.00442 & 0.00194 &  0.00387 & 0.00188 &  0.00412 & 0.00191 \\
 $regs$     & -0.00102 & 0.00167 & -0.00126 & 0.00157 & -0.00119 & 0.00163 \\
 $regw$     &  0.00270 & 0.00197 &  0.00208 & 0.00179 &  0.00226 & 0.00187 \\
 $marnev$   & -0.00496 & 0.00168 & -0.00475 & 0.00167 & -0.00508 & 0.00165 \\
 $marsdw$   & -0.00733 & 0.00216 & -0.00661 & 0.00217 & -0.00687 & 0.00212 \\
 $frelu6$   & -0.00234 & 0.00124 & -0.00207 & 0.00117 & -0.00233 & 0.00120 \\
 $frelu18$  &  0.00157 & 0.00069 &  0.00154 & 0.00067 &  0.00163 & 0.00067 \\
 $vm$       & -0.00865 & 0.00218 & -0.00821 & 0.00214 & -0.00842 & 0.00208 \\
 \hline\hline
 \end{tabular}
 \label{table:stats:CPS:q010}
 \end{table}

Table~\ref{t: cps q010 truncated } shows the acceptance rates,
inefficiencies, the equivalent sample sizes and the \ect{}s  of the
adaptive sampling schemes for the three prior distributions. The
adaptive independent \MH{} algorithms have lower inefficiencies than
the adaptive random walk Metropolis algorithms, and the copula based
proposals have in general the highest acceptance rates and lowest
inefficiencies.

\begin{table}[!ht]
 \centering
 \caption{Acceptance rates, inefficiencies (IF) based on the truncated kernel,
 equivalent sample size (ESS=10000/IF) and ECT = IF $\times$ time for 100 000 iterations
 for the 0.1-quantile regression applied to the current population survey data.}
 {\footnotesize
 \begin{tabular}{l | c | rrr | rrr | rrr}\hline\hline
 & & \multicolumn{3}{c|}{Inefficiency}  & \multicolumn{3}{c|}{Equivalent sample size}
 &\multicolumn{3}{c}{Timing}   \\\cline{3-11}
 Algorithm &  A. Rate & Min & Median & Max & Min & Median & Max & Min & Median & Max \\\hline
              & \multicolumn{10}{c}{Normal Prior}    \\\hline
 RWM          & 18.0 & 93.934 & 99.926 &102.211 &   98 &  100 &  106 & 2757 & 2933 & 3000 \\
 RWM3C        & 25.2 & 88.069 & 90.182 & 92.733 &  108 &  111 &  114 & 2598 & 2661 & 2736 \\
 IMH-MN-CL    & 47.3 &  3.855 &  4.321 &  4.957 & 2017 & 2315 & 2594 &  128 &  143 &  164 \\
 IMH-TCT-CL   & 54.2 &  3.353 &  3.609 &  4.049 & 2470 & 2770 & 2982 &  849 &  914 & 1025 \\
 IMH-TCT-CL-A & 51.7 &  3.125 &  3.642 &  4.376 & 2285 & 2746 & 3200 &  889 & 1036 & 1245 \\
 \hline       & \multicolumn{10}{c}{Double exponential Prior}    \\\hline
 RWM          & 15.5 &102.917 &106.161 &108.036 &   93 &   94 &   97 & 3551 & 3662 & 3727 \\
 RWM3C        & 22.4 & 93.033 & 95.986 & 98.044 &  102 &  104 &  107 & 3227 & 3329 & 3401 \\
 IMH-MN-CL    & 41.6 &  4.971 &  5.452 & 16.263 &  615 & 1834 & 2012 &  188 &  206 &  614 \\
 IMH-TCT-CL   & 47.8 &  4.273 &  4.568 &  4.993 & 2003 & 2189 & 2340 & 1452 & 1552 & 1697 \\
 IMH-TCT-CL-A & 46.3 &  4.166 &  4.623 &  5.859 & 1707 & 2163 & 2401 & 1732 & 1923 & 2437 \\
 \hline       & \multicolumn{10}{c}{Mixture of Normals Prior}    \\\hline
 RWM          & 22.8 & 81.008 & 83.060 & 92.968 &  108 &  120 &  123 & 3006 & 3083 & 3450 \\
 RWM3C        & 23.2 & 84.279 & 86.135 & 92.548 &  108 &  116 &  119 & 3157 & 3226 & 3466 \\
 IMH-MN-CL    & 46.0 &  4.227 &  4.688 &  8.217 & 1217 & 2133 & 2366 &  205 &  227 &  398 \\
 IMH-TCT-CL   & 54.4 &  3.507 &  3.722 &  6.412 & 1560 & 2687 & 2852 & 2066 & 2193 & 3777 \\
 IMH-TCT-CL-A & 49.7 &  3.938 &  4.365 &  9.977 & 1002 & 2291 & 2540 & 4939 & 5475 &12514 \\
 \hline\hline
 \end{tabular}
 }
  \label{t: cps q010 truncated }
 \end{table}

 \section{Binary random effects model}\label{section:adapt:import:sampl}
This section considers adaptive sampling in a binary random effects
model where the random effects are integrated out using importance
sampling. However, the same ideas can be applied to other  random
effects models.

Suppose there are $N$ groups with $J_i$ observations in the $i$th group, such that the probability of the $ij$th binary
 response is given by the probit model,
 \begin{align}
 \label{eq:probit:ran:eff:mod}
 \Pr(y_{ij} = 1 \mid \mu_{i}, \beta,x_{ij} ) &= \Phi(\mu_{i} + x_{ij}^T\beta),\quad \mu_{i} &\sim N(0,\sigma_\mu^2)
 \quad i=1, \dots, N , j=1,\dots, J_i\ ,
 \end{align}
and $\Phi(\cdot)$ is the standard normal \cdf. Let $ \bm{\theta}= (\beta, \sigma_\mu^2)$ be the parameter vector, $\bfyi = (y_{i1},\dots, y_{iJ_i}), \bfxi = (x_{i1},\dots, x_{iJ_i})$ are the vectors of observations on the $i$th group,
$\bfy$ is the vector of all the observations and $\bfx$ is the vector of all the covariates. Then
\begin{align}
 \label{eq:lik}
 p(\bfy \mid \bftheta ,\bfx ) &=  \prod_{i=1}^N p(\bfyi\mid \bftheta ,\bfxi ) \ , \quad p(\bfyi  \mid \bftheta , \bfxi  ) = \int p( \bfyi\mid \mu_i, \bftheta , \bfxi )p(\mu_i \mid  \bftheta ) d\mu_i
 \end{align}
 We form proposals for the posterior $p(\bftheta\mid \bfy , \bfx)$
 with the random effects integrated out because in many applications there are too
 many random effects to include in the adaptation.
 We use importance sampling to integrate out the random effects with the
 importance density based on previous iterates. Let $\Ehat(\mu_i\mid \bfy)$ and
 $\varhat(\mu_i\mid \bfy)$ be estimates of the posterior mean and variance of $\mu_i$. We use the importance density $h(\mu_i)\sim N\biggl( \Ehat(\mu_i\mid \bfy),\kappa \varhat(\mu_i\mid \bfy)\biggr )  $ to efficiently
integrate out $\mu_i$ in \eqref{eq:lik}, with $\kappa =4 $,  using
\begin{align} \label{e: phat}
 p(\bfyi \mid \bftheta ,  \bfxi) & = \int \biggl ( p( \bfyi \mid \mu_i, \bftheta , \bfxi )p(\mu_i \mid  \bftheta )/h(\mu_i)\biggr )  h(\mu_i) d\mu_i\nonumber \\
& \approx \frac{1}{M}\sum_{j=1}^{M}p( \bfyi\mid \mu_i^{[j]}, \bftheta , \bfxi )p(\mu_i^{[j]} \mid  \bftheta )/h(\mu_i^{[j]})\ ,
\end{align}
where the $\mu_i^{[j]}$ are generated from $h(\mu_i)$. We obtain $\Ehat(\mu_i\mid \bfy)$ by first writing
$E(\mu_i\mid \bfy) = E(E(\mu_i\mid \bfyi,\bftheta)\mid \bfy)$ , so that
\begin{align}
\Ehat(\mu_i\mid \bfy) & \approx \frac{1}{L}\sum_{l=1}^{L}E(\mu_i\mid \bfy,\bftheta^{[l]})\ , \\
E(\mu_i\mid \bfyi,\bftheta^{[l]}) & = \frac{1}{p(\bfyi \mid \bftheta^{[l]}, \bfxi  )}
\int \biggl ( \mu_i p(\bfyi \mid \bftheta^{[l]} , \bfxi )p(\mu_i \mid  \bftheta^{[l]}) /h(\mu_i) \biggr ) h(\mu_i) d\mu_i\nonumber \\
& \approx \frac{1}{\phat ( \bfyi \mid \bftheta^{[l]}, \bfxi  )}\frac{1}{M}\sum_{j=1}^{M}
\mu_i^{[j]} p(\bfyi\mid \mu_i^{[j]}, \bftheta^{[l]} , \bfxi )p(\mu_i^{[j]} \mid  \bftheta^{[l]} )/h(\mu_i^{[j]})
\end{align}
where the $\bftheta^{[l]}$ are iterates from the adaptive sampling
and $\phat (\bfyi \mid \bftheta^{[l]},  \bfxi )$ is constructed by
the right side of \eqref{e: phat}. The estimate $\Ehat(\mu_i^2\mid
\bfy)$ is obtained similarly and $\varhat(\mu_i\mid \bfy)$ is then
computed as $\Ehat(\mu_i^2\mid \bfy) - \biggl ( \Ehat(\mu_i\mid
\bfy)  \biggr )^2$.

We update the importance density after every $L$ accepted values of the adaptive sampling scheme, with $L$ given in
appendix~\ref{S:sim details}

\section*{Pap smear data}
We applied the probit random effects model
\eqref{eq:probit:ran:eff:mod} to data collected in a discrete choice
experiment designed to study factors that may determine whether a woman
chooses to have a Pap smear test to detect cervical cancer. The
study is described by  \cite{fieb:hall:2005} and is based on  $N=79$
Australian women, where each woman was presented with $J_i=32$
different scenarios and for each scenario asked whether she would
choose to have a Pap smear test. Thus, there are 32 repeated binary
observations for each woman. Table~\ref{t: pap smear variables}
lists the covariates in the study.

 \begin{table}[!ht]
 \centering
 \caption{Covariates for the Pap smear data set}
 \begin{tabular}{l l}\hline
 $knowgp:$  & 1 if the GP is known to the patient and 0 otherwise;\\
 $sexgp:$   & 1 if the GP is male and 0 otherwise;\\
 $testdue:$ & 1 if the patient is due or overdue for a Pap test and 0 otherwise; and\\
 $drrec:$   & 1 if GP recommends that the patient has a Pap test and 0 otherwise.\\
 $papcost:$ & Cost of the Pap test in AU\$
 (2 levels).\\\hline
 \end{tabular}
 \label{t: pap smear variables}
 \end{table}
We fitted the binary random effects model with 7 parameters and 79 random effects to the data with
an ${\IG}(0.01,0.01)$ prior for $\sigma_\mu^2$. For the double exponential prior for $\beta$, the prior for $\tau$ is ${\IG}(0.01,0.01)$. For the mixture of normals prior, we set $\tau_S^2=0.0001$ and $\tau_L^2=10000$.
 and the prior for $\omega$ is uniform on (0,1). In the adaptive sampling scheme we generated  $\log \sigma_{\mu}^2$ because it was unconstrained.
The initial values and proposal distributions for the adaptive independent \MH{} algorithms were obtained by running the 3 component adaptive random walk for 2 000 iterations.
To initialize all the adaptive random walk  algorithms we first used the \MATLAB{} function
\verb"glmfit" to estimate the regression coefficients and their standard
errors with the random effects set identically to zero. To integrate out the random effects in the
adaptive random walk proposals we began with the proposal importance density $h(\mu_i)\sim
N(0,1.5\sigma^2)$, with $\sigma^2=1$ initially.

Table~\ref{table:stats:Paptest} summarizes the posterior distributions of the parameters.
\begin{table}[!ht]
 \centering
 \caption{Summary of statistics of the posterior distribution of the probit random effects model
 for the Pap test data under the normal, double exponential (with $\theta = \log (\tau$)) and mixture of normals (with $\theta = \text{logit} (\omega) $ )  prior distributions. }
 \begin{tabular}{l | cc|cc|cc}\hline\hline
 Parameter            & \multicolumn{2}{c|}{Normals}&\multicolumn{2}{c|}{Double Exponential}&
 \multicolumn{2}{c}{Mixture of Normals}\\\cline{2-7}
                      & Mean    & S. Dev. &  Mean   & S. Dev. & Mean    & S. Dev. \\\hline
 $\theta$             &     -   &    -    & -0.5683 &  0.4488 & -1.0691 &  0.8988 \\
 $\log(\sigma_\mu^2)$ &  0.5792 &  0.2081 &  0.5598 &  0.2151 &  0.5680 &  0.1976 \\
 constant             &  0.2914 &  0.1797 &  0.2646 &  0.1865 &  0.2808 &  0.1833 \\
 $knowgp$             & -0.3052 &  0.0633 & -0.2882 &  0.0658 & -0.2846 &  0.0934 \\
 $sexgp$              &  0.6582 &  0.0656 &  0.6483 &  0.0677 &  0.6591 &  0.0690 \\
 $testdue$            & -1.1794 &  0.0750 & -1.1688 &  0.0807 & -1.1769 &  0.0773 \\
 $drrec$              & -0.4978 &  0.0737 & -0.4772 &  0.0739 & -0.4941 &  0.0725 \\
 $papcost$            &  0.0091 &  0.0028 &  0.0091 &  0.0030 &  0.0085 &  0.0028 \\
 \hline\hline
 \end{tabular}
 \label{table:stats:Paptest}
 \end{table}

Table \ref{t: pap truncated}  shows the acceptance
rates, inefficiencies,  equivalent sample sizes, and \ect{}s for each
algorithm.  The copula based sampling schemes have the highest
acceptance rates and the smallest inefficiency factors, with the antithetic proposal being the best for the normal and double exponential priors, where the acceptance rates are at least 70\%.

 \begin{table}[!ht]
 \centering
 \caption{Acceptance rates, inefficiencies (IF) based on the truncated kernel,
 equivalent sample size (ESS=10000/IF) and ECT = IF $\times$ time for 100 000 iterations
 for the probit random effects applied to the Pap test data.}
 {\footnotesize
 \begin{tabular}{l | c | rrr | rrr | rrr}\hline\hline
 & & \multicolumn{3}{c|}{Inefficiency}  & \multicolumn{3}{c|}{Equivalent sample size}
 &\multicolumn{3}{c}{ECT}   \\\cline{3-11}
 Algorithm &  A. Rate & Min & Median & Max & Min & Median & Max & Min & Median & Max \\\hline
              & \multicolumn{10}{c}{Normal Prior}    \\\hline
 RWM          & 28.6 & 18.210 &  23.039 &  32.306 &  310 &  434 &  549 & 618517 &  782509 &  1097268 \\
 RWM3C        & 28.8 & 22.623 & 24.766 & 33.667 &  297 &  404 &  442 & 750806 & 821899 & 1117299 \\
 IMH-MN-CL    & 62.1 &  2.356 &  2.590 &  2.999 & 3334 & 3862 & 4244 &  77641 &  85324 &   98824 \\
 IMH-TCT-CL   & 71.6 &  1.803 &  2.086 &  2.226 & 4493 & 4793 & 5548 &  59834 &  69253 &   73879 \\
 IMH-TCT-CL-A & 70.9 &  0.968 &  1.453 &  1.893 & 5282 & 6884 &10329 &  32002 &  48017 &   62580 \\
 \hline       & \multicolumn{10}{c}{Double exponential Prior}    \\\hline
 RWM          & 26.8 & 25.337 & 30.855 & 39.939 &  250 &  325 &  395 & 833790 & 1015358 & 1314306 \\
 RWM3C        & 26.8 & 24.894 & 30.907 & 36.229 &  276 &  324 &  402 & 821509 & 1019936 & 1195573 \\
 IMH-MN-CL    & 57.9 &  2.656 &  2.860 &  3.542 & 2823 & 3502 & 3765 &  86895 &   93583 &  115885 \\
 IMH-TCT-CL   & 71.8 &  1.922 &  2.151 &  2.514 & 3978 & 4650 & 5202 &  63124 &   70627 &   82551 \\
 IMH-TCT-CL-A & 70.7 &  0.974 &  1.526 &  2.209 & 4527 & 6555 &10272 &  32187 &   50462 &   73036 \\
 \hline       & \multicolumn{10}{c}{Mixture of Normals Prior}    \\\hline
 RWM          & 25.4 & 27.979 & 32.678 & 90.248 &  111 &  307 &  357 & 1014949 & 1185408 & 3273811 \\
 RWM3C        & 24.2 & 26.739 & 34.183 & 52.536 &  190 &  293 &  374 & 1004583 & 1284255 & 1973767 \\
 IMH-MN-CL    & 58.4 &  2.369 &  3.058 &  3.720 & 2688 & 3271 & 4221 &   82817 &  106888 &  130032 \\
 IMH-TCT-CL   & 71.5 &  2.011 &  2.244 &  3.364 & 2973 & 4462 & 4971 &   73396 &   81889 &  122746 \\
 IMH-TCT-CL-A & 62.2 &  2.365 &  2.635 &  6.221 & 1607 & 3794 & 4229 &   88108 &   98204 &  231813 \\
 \hline\hline
 \end{tabular}
 }
  \label{t: pap truncated}
 \end{table}

\section{Summary}
Our article proposes a new copula based adaptive sampling scheme and
a generalization of the two component adaptive random walk designed
to explore the target space more efficiently than the proposal of
\cite{roberts06_applied}. We studied  the performance of these
sampling schemes as well as the adaptive independent \MH{} sampling
scheme proposed by \cite{gior:kohn:2008} which is based on a mixture
of normals. All the sampling schemes performed reliably on the
examples  studied in the article, but we found that the adaptive
independent \MH{} schemes had inefficiency factors that were often
much lower and acceptance rates that were much higher than the
adaptive random walk schemes. The copula based adaptive scheme often
had the smallest inefficiency factors and highest acceptance rates.
For acceptance rates over 70\% the antithetic version of the copula based
approach was the most efficient.
Our results suggest that the copula based proposal provides  an
attractive approach to adaptive sampling, especially for higher dimensions.
However, the mixture of normals approach of
\cite{gior:kohn:2008} also performed well and is useful for
complicated and possibly multimodal posterior distributions.

\section*{Acknowledgment}
The research of Robert Kohn, Ralph Silva and Xiuyan Mun was partially supported by an ARC Discovery
Grant DP0667069. We thank Professor Garry Barret for  the CPS data and Professor Denzil Fiebig for
the Pap smear data.
%\begin{comment}
%\begin{appendices}
\begin{appendix}
\section{Details of the Simulation} \label{S:sim details}
All the computations were done on Intel Core 2 Quad 2.6 Ghz
processors, with 4GB RAM (800Mhz) on a GNU/Linux platform using
\MATLAB{} 2007b. However, in the TCT algorithm
we computed the univariate \cdfs{} and inverse \cdfs{} of the $t$,
normal and mixture of normals distributions using \MATLAB{} mex files
based on the corresponding \MATLAB{} code. In addition, to speed up the
computation, we tested each marginal for normality using the
Jarque-Bera test at the 5\% level. If normality was not rejected
then we fitted a normal density to the marginal. Otherwise, we
estimated the marginal density by a mixture of normals.

In Stage 1 of the adaptive sampling schemes IMH-MN-CL and IMH-MN-SA
that use  a multivariate mixture of normals, the number of
components ($nc$) used in the third term $g_3(x)$ of the mixture
\eqref{eq:aimh proposal} is determined by the dimension of the
parameter vector ($dim$) and the number of accepted draws ($accep$)
to that stage of the simulation. In particular, $nc = 1$ if
$accep/dim < 40$, $nc = 2$ if $40 \le accep/dim <100$,  $nc = 3$ if
$100 \le accep/dim <200$ and $nc = 4$ if $ accep/dim \ge 200$.

We now give details of the number of iterations, burn-in and
updating schedules for all the adaptive independent \MH{} schemes in
the paper. In addition, we update the proposal in Stage 1 if in 100
successive iterations the acceptance rate is lower than 0.01.

\noindent
 {\sf Logistic regression, HMDA data}
      \begin{itemize}
     \item  Normal prior:      end of first stage = 5 000;
     burn-in = 75 000;
     number of iterations: 100 000;
     updates = [50, 100, 150, 200, 300, 500, 700, 1000, 2000, 5000, 10000, 20000, 30000, 50000, 75000].
     \item Double exponential prior:
     end of first stage = 5 000;
     burn-in = 100 000;
     number of iterations: 150 000;
     updates = [50, 100, 150, 200, 300, 500, 700, 1000, 2000, 5000, 10000, 20000, 30000, 50000, 75000, 100000].
     \item  Mixture of normals prior:
     end of first stage = 100 000;
     burn-in = 300 000;
     number of iterations: 400 000;
     updates = [100, 150, 200, 300, 500, 700, 1000, 2000, 3000, 5000, 7500, 10000, 15000, 20000, 30000, 50000, 75000, 100000, 125000, 150000, 175000, 200000, 225000, 250000, 300000].
     \end{itemize}
\begin{comment}
    \noindent
    {\sf Logistic regression, Labor force participation data}
     \begin{itemize}
     \item Normal prior:
     end of first stage = 3 000;
     burn-in = 75 000;
     number of iterations: 100 000;
     updates = [50, 100, 150, 200, 300, 500, 700, 1000, 2000, 5000, 10000, 20000, 30000, 50000, 75000];
     \item Double exponential prior:
     end of first stage = 3 000;
     burn-in = 100 000;
     number of iterations: 150 000;
     updates = [50, 100, 150, 200, 300, 500, 700, 1000, 2000, 5000, 10000, 20000, 30000, 50000, 75000, 100000].
     \item Mixture of normals prior:
     end of first stage = 100 000;
     burn-in = 150 000;
     number of iterations: 200 000;
     updates = [100, 150, 200, 300, 500, 700, 1000, 2000, 3000, 5000, 7500, 10000, 15000, 20000, 30000, 50000, 75000, 100000, 150000].
     \end{itemize}
\end{comment}
\noindent
     {\sf Bayesian quantile regression, CPS data}.
      \begin{itemize}
     \item
     Normal and double exponential priors, quantiles 0.1, 0.5 and 0.9:
     end of first stage = 3 000;
     burn-in = 150 000;
     number of iterations: 200 000;
     updates = [100, 150, 200, 300, 500, 700, 1000, 2000, 3000, 5000, 7500, 10000, 15000, 20000, 30000, 50000, 75000, 100000, 150000].
     \item Mixture of normals prior, quantiles 0.1, 0.5 and 0.9:
     end of first stage = 200 000;
     burn-in = 400 000;
     number of iterations: 500 000;
     updates = [100, 150, 200, 300, 500, 700, 1000, 2000, 3000, 5000, 7500, 10000, 15000, 20000, 30000, 50000, 75000, 100000, 125000, 150000, 175000, 200000, 225000, 250000, 275000, 300000, 325000, 350000, 375000, 400000].
 \end{itemize}

 \noindent
 {\sf Probit random effects model, Pap smear data}. For all three priors,
%  \begin{itemize}
%    \item Normal prior:
     end of first stage = 5 000;
     burn-in = 10 000;
     number of iterations: 20 000;
     updates = [20, 50, 100, 150, 200, 300, 400, 500, 600, 700, 800, 900, 1000, 1100, 1200, 1300, 1400, 1500, 2000, 2500, 3000, 3500, 4000, 4500, 5000, 6000, 7000, 8000, 9000, 10000, 12000, 15000].
%     \end{itemize}

In this example the importance sampling density is updated every $L
= 100$ iterations.

We now give the details of the sampling for both adaptive random
walk Metropolis algorithms. For the HMDA  data the number of burn-in
iterations was 300 000 and the total number of iterations was 500
000 for all three priors. The corresponding numbers for the CPS data
with normal and double exponential priors are 500 000 and 1000 000,
and for the mixture of normals prior 1000 000 and 1500 000. The
corresponding numbers for the Pap smear data are 30 000 and 50 000.
%\end{appendices}
\end{appendix}

%\bibliography{mixtures}
 
\end{document}